\documentclass[conference]{IEEEtran}
\usepackage{algorithm,algorithmic}
\makeatletter
\newcommand\fs@betterruled{%
  \def\@fs@cfont{\bfseries}\let\@fs@capt\floatc@ruled
  \def\@fs@pre{\vspace*{5pt}\hrule height.8pt depth0pt \kern2pt}%
  \def\@fs@post{\kern2pt\hrule\relax}%
  \def\@fs@mid{\kern2pt\hrule\kern2pt}%
  \let\@fs@iftopcapt\iftrue}
\floatstyle{betterruled}
\restylefloat{algorithm}
\makeatother
\newcommand{\squeezeup}{\vspace{-4mm}}

\newcommand{\squeezeupann}{\vspace{-1mm}}

\usepackage{cite}
\usepackage{amsmath,amssymb,amsfonts}
\usepackage{algorithmic}
\usepackage{graphicx}
\usepackage{textcomp}
\usepackage{comment}
\usepackage{enumitem}

\def\BibTeX{{\rm B\kern-.05em{\sc i\kern-.025em b}\kern-.08em
    T\kern-.1667em\lower.7ex\hbox{E}\kern-.125emX}}
\begin{document}

\title{Simultaneous Data Communication and Channel Estimation in Multi-User Full Duplex MIMO Systems}

\author{\IEEEauthorblockN{Md Atiqul Islam\IEEEauthorrefmark{2}, George C. Alexandropoulos\IEEEauthorrefmark{4}, and Besma Smida\IEEEauthorrefmark{2}}
\IEEEauthorblockA{{\IEEEauthorrefmark{2}Department of Electrical and Computer Engineering, University of Illinois at Chicago, USA}\\
\IEEEauthorrefmark{4}Department of Informatics and Telecommunications, National and Kapodistrian University of Athens, Greece\\
emails: \{mislam23,smida\}@uic.edu, alexandg@di.uoa.gr
}}
\maketitle

\begin{abstract}
In this paper, we study Simultaneous Communication of Data and Control (SCDC) information signals in Full Duplex (FD) Multiple-Input Multiple-Output (MIMO) wireless systems. In particular, considering a FD MIMO base station serving multiple single-antenna FD users, a novel multi-user communication scheme for simultaneous DownLink (DL) beamformed data transmission and UpLink (UL) pilot-assisted channel estimation is presented. Capitalizing on a recent FD MIMO hardware architecture with reduced complexity self-interference analog cancellation, we jointly design the base station’s transmit and receive beamforming matrices as well as the settings for the multiple analog taps and the digital SI canceller with the objective to maximize the DL sum rate. Our simulation results showcase that the proposed approach outperforms its conventional half duplex counterpart with $50\%$ reduction in hardware complexity compared to the latest FD-based SCDC schemes. 
\end{abstract}


\section{Introduction}
Multi-user Full Duplex (FD) Multiple-Input Multiple-Output (MIMO) communication technology has the potential of substantial spectral efficiency improvement and simplification of the control information exchange over conventional frequency- and time-division duplexing systems through concurrent UpLink (UL) and DownLink (DL) communication in the same frequency and time resources \cite{sabharwal2014band,duarte2012experiment,bharadia2013full,smida2017reflectfx,islam2019comprehensive,alexandropoulos2020full,islam2019unified,riihonen2011mitigation,everett2016softnull}. 

The main bottleneck of FD MIMO systems is the in-band Self Interference (SI) signal at the reception side resulting from simultaneous transmission and reception, as well as limited Transmitter (TX) and Receiver (RX) isolation. Therefore, a combination of propagation domain isolation, analog domain suppression, and digital SI cancellation techniques are adopted in practice to suppress the strong SI below the noise floor. In FD MIMO systems, the suppression techniques are particularly challenging due to higher SI components, as a consequence of the increased number of transceiver antennas. Analog SI cancellation in FD MIMO systems can be implemented by subtracting a processed copy of every TX signal from every RX input to avoid saturation of the RXs' Radio Frequency (RF) chain~\cite{sabharwal2014band}. However, the hardware requirements for such an approach scale with the number of TX/RX antennas, rendering the implementation of analog SI a core design bottleneck. In \cite{riihonen2011mitigation,everett2016softnull}, authors presented spatial suppression techniques that alleviate the need of analog SI cancellation, relying solely on digital TX/RX beamforming. In \cite{alexandropoulos2017joint}, a joint design of multi-tap analog cancellation and TX/RX beamforming, where the number of taps does not scale with the product of TX and RX antenna elements, was proposed. According to this work, every analog tap refers to a line of fixed delay, variable phase shifter, and attenuator. The authors in \cite{islam2019unified} presented a unified beamforming approach including both Analog and Digital (A/D) SI cancellation that further improves the achievable rate performance under practical transceiver imperfections.

Recently, in \cite{du2015mu,du2016sequential,mirza2018performance}, simultaneous DL data transmission and UL Channel State Information (CSI) reception has been considered at an FD MIMO Base Station (BS) serving multiple Half Duplex (HD) User Equipment (UE) nodes. In these studies, the UEs transmit training symbols through the UL channel in a Time Division Multiple Access (TDMA) manner, which are utilized by the BS to estimate the DL channels leveraging channel reciprocity, while at the same time transmitting the DL payload to UEs for whom the DL CSI is already available. However, the adopted multi-user models consider perfect analog cancellation that is based on conventional FD MIMO architectures with fully connected analog cancellation, interconnecting all TX antenna elements in the FD node with all its RX antennas. In \cite{islam2020simultaneous}, a low complexity solution of simultaneous DL data transmission and UL CSI acquisition was proposed for a single user FD MIMO system.

\begin{figure*}[!t]
\centering
\includegraphics[width=0.95\textwidth]{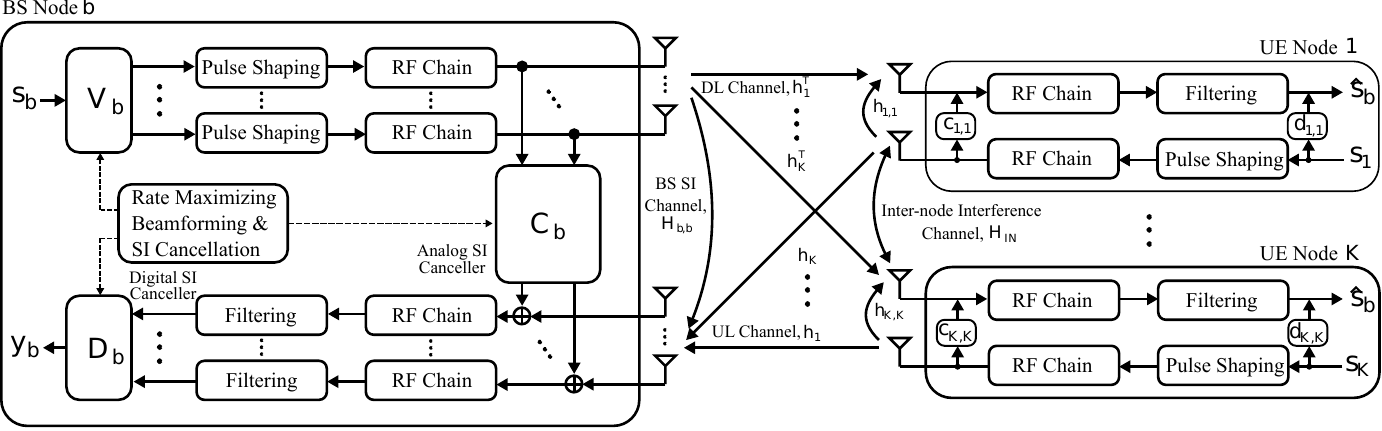}
\caption{The considered multi-user FD MIMO system model for simultaneous DL data transmission and UL pilot-assisted channel estimation. BS node $b$ and all UE nodes deploy A/D SI cancellation with the former node $b$ realizing \cite{islam2019unified}'s reduced complexity multi-tap analog cancellation.}
\squeezeup
\label{fig: transmission}
\end{figure*}

In this paper, we present a multi-user MIMO communication system for Simultaneous Communication of Data and Control (SCDC) information signals, capitalizing on the FD MIMO architecture of \cite{islam2019unified} combining TX beamforming with A/D SI cancellation. Exploiting channel reciprocity and relying on FD operation, the proposed system performs joint digital TX beamforming for DL communication and UL pilot-assisted CSI estimation. Considering realistic modeling for imperfect channel estimation and CSI delay error, we present a joint optimization framework for the DL rate optimization and the accurate CSI estimation. Our simulation results showcase superior achievable DL rate performance for the proposed SCDC scheme with more than $50\%$ reduction in the analog cancellation hardware complexity compared to the conventional FD MIMO architectures with fully connected analog cancellation.
 
\textit{Notation:} Vectors and matrices are denoted by boldface lowercase and boldface capital letters, respectively. The transpose, Hermitian transpose, and conjugate of $\mathbf{A}$ are denoted by $\mathbf{A}^{\rm T}$, $\mathbf{A}^{\rm H}$, and $\mathbf{A}^*$, respectively, and $\det(\mathbf{A})$ is $\mathbf{A}$'s determinant, while $\mathbf{0}_{m\times n}$ ($m\geq2$ and $n\geq1$) represents the $m\times n$ matrix with all zeros. $\|\mathbf{a}\|$ stands for the Euclidean norm of $\mathbf{a}$. $[\mathbf{A}]_{i,j}$, $[\mathbf{A}]_{(i,:)}$, and $[\mathbf{A}]_{(:,j)}$ represent $\mathbf{A}$'s $(i,j)$-th element, $i$-th row, and $j$-th column, respectively, while $[\mathbf{a}]_{i}$ denotes the $i$-th element of $\mathbf{a}$. $\text{sort}(\mathbf{A})$ represents a rearranged row formation of $\mathbf{A}$, where the subsequent rows have decreasing Euclidean norm.  $\mathbb{C}$ represents the complex number set, $\mathbb{E}\{\cdot\}$ is the expectation operator, and $|\cdot|$ denotes the amplitude of a complex number. $x\sim\mathcal{CN}(0,\sigma^2)$ represents a circularly symmetric complex Gaussian random variable with zero mean and variance $\sigma^2$.

\section{System and Signal Models}
A multi-user MIMO communication is considered with a BS node $b$ containing $N_b$ TX and RX antennas, and $K$ single-antenna FD UE nodes, as shown in Fig. \ref{fig: transmission}. Each antenna is attached to a dedicated TX/RX RF chain at all the nodes. A multi-tap analog SI canceller is applied in the FD BS node $b$, whereas each UE node deploys a single-tap SI canceller. The FD BS node is capable of performing digital TX beamforming realized, for simplicity, with linear filters. We consider UL/DL channel reciprocity and focus on simultaneous data communication and channel estimation, where the DL is intended for information data communication while the UL is used for transmitting training signals from UEs to the BS. The UL CSI estimation is used for designing the DL precoder to digitally process the data signals before transmission.

We assume that, for every channel use, the BS node $b$ transmits the complex-valued information data symbols $\mathbf{s}_b\in \mathbb{C}^{m_b\times 1}$ (chosen from a discrete modulation set) using the unit norm digital precoding vector $\mathbf{V}_b\in\mathbb{C}^{N_b\times m_b}$, where the $m_b$ is the number of data streams. In the UL direction, the $K$ users simultaneously send orthogonal training symbols to the BS node $b$, which can be denoted as $\mathbf{s}_K \in \mathbb{C}^{K\times 1}$.  The signal transmissions at both BS and UE nodes are power limited to ${P}_b$ and ${P}_k$, respectively. Specifically, the DL signal is such that $\mathbb{E}\{\|\mathbf{V}_b\mathbf{s}_b\|^2\}\leq {P}_b$, whereas the UL signal from all $K$ UEs is constrained as $\mathbb{E}\{\|\mathbf{s}_K\|^2\}\leq {P}_K$. 

\subsection{Channel Modeling}
The Rayleigh faded DL channel from BS node $b$ to UE node $k$, denoted by $\mathbf{h}_{k}^{\rm T}\in \mathbb{C}^{1\times N_b}$, is modeled as Independent and Identically Distributed (IID) $\mathcal{C}\mathcal{N}(0,l_{K})$, where $l_{K}$ is the UL pathloss. Stacking all UE channel vectors, the DL channel for $K$ users can be expressed as $\mathbf{H}= [\mathbf{h}_{1}\,\mathbf{h}_{2}\, \cdots\, \mathbf{h}_{K}]^{\rm T}\in\mathbb{C}^{K\times N_b}$. Based on the UL/DL channel reciprocity, the Rayleigh faded UL channel is $\mathbf{H}^{\rm T}\in\mathbb{C}^{N_b\times K}$. As all nodes are capable of FD operation, the simultaneous DL data and UL training transmission induce SI in the RXs of the BS and UEs, respectively. We consider the Rician fading model for the SI channels denoted by $\mathbf{H}_{b,b} \in \mathbb{C}^{N_b\times N_b}$ for the BS and ${h}_{k,k}\in\mathbb{C},\,\forall k = 1,2,\ldots,K$, for the UEs with Rician factor $\kappa$ and pathlosses $l_{b,b}$ and $l_{k,k}$ at nodes $b$ and $k$, respectively \cite{duarte2012experiment}. Using matrix notation, the SI channel for $K$ users can be expressed as $\mathbf{H}_{K,K}=\text{diag}\{{h}_{1,1},{h}_{2,2},\cdots,{h}_{K,K}\} \in \mathbb{C}^{K\times K}$. Similarly, the inter-node interference channel between $K$ users containing only off-diagonal elements is denoted as $\mathbf{H}_{\text{IN}} \in \mathbb{C}^{K\times K}~\mathcal{C}\mathcal{N}(0,l_{\text{IN}})$, where $l_{\text{IN}}$ is the inter-node interference channel pathloss.

\subsection{Signal Modeling at the BS and UE Nodes}
Because of FD operation, BS node $b$ RXs receive the training symbols transmitted from $K$ users as well as the SI signal induced by simultaneous data transmission. As shown in Fig. \ref{fig: transmission}, the SI is suppressed using low complexity analog SI cancellation, which is followed by a digital canceller in the baseband of BS node $b$. Denoting analog and digital SI cancellers as $\mathbf{C}_{b}\in\mathbb{C}^{N_b\times N_b}$ and $\mathbf{D}_{b}\in\mathbb{C}^{N_b\times N_b}$, respectively, the received baseband signal at BS node $b$, $\mathbf{y}_{b}\in\mathbb{C}^{N_b\times 1}$, can be expressed as
\begin{equation}
    \begin{split}
        \mathbf{y}_{b} &\triangleq \mathbf{H}^{\rm T}\mathbf{s}_K + \left(\mathbf{H}_{b,b}+\mathbf{C}_{b}+\mathbf{D}_{b}\right)\mathbf{V}_b\mathbf{s}_b + \mathbf{n}_b,\\
        &= \mathbf{H}^{\rm T}\mathbf{s}_K + \widetilde{\mathbf{H}}_{b,b}\mathbf{V}_b\mathbf{s}_b + \mathbf{n}_b,
    \end{split}
\end{equation}
where $\mathbf{n}_{b}\in \mathbb{C}^{N_b\times 1}$ is the zero-mean Additive White Gaussian Noise (AWGN) with variance $\sigma^2_{b}\mathbf{I}_{N_b}$. Here, $\widetilde{\mathbf{H}}_{b,b}\triangleq\left(\mathbf{H}_{b,b}+\mathbf{C}_{b}+\mathbf{D}_{b}\right)$ represents the residual SI channel after A/D SI cancellation. 

Similarly, after A/D SI cancellation at each UE node, the precoded DL signal received at $K$ user nodes, $\mathbf{y}_{K}\in\mathbb{C}^{K\times 1}$ is written as
\begin{align}
    \nonumber\mathbf{y}_{K} &\triangleq \mathbf{H}\mathbf{V}_b\mathbf{s}_b + \left(\mathbf{H}_{K,K}+\mathbf{C}_{K}+\mathbf{D}_{K}\right)\mathbf{s}_{K} + \mathbf{H}_{\text{IN}}\mathbf{s}_{K} + \mathbf{n}_{K},\\
        &= \mathbf{H}\mathbf{V}_b\mathbf{s}_b+ \widetilde{\mathbf{H}}_{K,K}\mathbf{s}_{K}+ \mathbf{H}_{\text{IN}}\mathbf{s}_{K} + \mathbf{n}_{K},
\end{align}
where $\mathbf{n}_{K}\in \mathbb{C}^{K\times 1}$ is the zero-mean AWGN with variance $\sigma^2_{K}\mathbf{I}_{K}$. Here, $\mathbf{C}_{K}\triangleq\text{diag}\{{c}_{1,1},{c}_{2,2},\cdots,{c}_{K,K}\} \in \mathbb{C}^{K\times K}$, $\mathbf{D}_{K}\triangleq\text{diag}\{{d}_{1,1},{d}_{2,2},\cdots,{d}_{K,K}\} \in \mathbb{C}^{K\times K}$, where the diagonal elements of $\mathbf{C}_{K}$ and $\mathbf{D}_{K}$ represent the analog and digital SI canceller of each UE node, and $\widetilde{\mathbf{H}}_{K,K}\triangleq\left(\mathbf{H}_{K,K}+\mathbf{C}_{K}+\mathbf{D}_{K}\right)$ is the residual SI channel at all UE nodes. 

\section{Proposed SCDC Scheme}\label{sec: SDCD}
In this section, we present a multi-user FD MIMO based  data communication and channel estimation scheme. First, we describe the CSI acquisition and DL beamforming scheme in Section \ref{ssec: Proposed Strategy}. Then, in Section \ref{ssec: CSI Error}, we characterize the realistic channel estimation error, CSI delay error, and average achievable DL rate.

\subsection{Proposed CSI Acquisition and Beamforming}\label{ssec: Proposed Strategy}
\begin{figure}
    \centering
    \includegraphics[width=\linewidth]{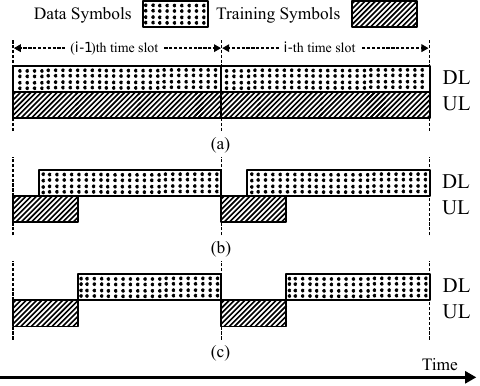}
    \caption{(a) Proposed SCDC Scheme, (b) Sequential Beamforming with DL data transmission and UL CSI, and (c) Conventional HD Beamforming with UL training over two consecutive time slots.}
    \label{fig: trans_protocol}
\end{figure}

The proposed SCDC scheme is illustrated in Fig. \ref{fig: trans_protocol}(a), where DL is dedicated for data transmission and UL is accessed to transmit training symbols.
We assume a time division duplexing based data transmission, where the considered channels remains constant for all the channel uses in a time slot and the channels of successive time slots are temporally correlated\cite{kobayashi2011training,tse2005fundamentals}. Each time slot contains $T$ symbols (i.e. $T$ channel uses). The proposed data transmission scheme and CSI acquisition scheme is described as follows:

\begin{enumerate}
    \item For data transmission at any $i$th time slot, first, we obtain the UL channel estimate of $(i\!-\!1)$th time slot using $T$ orthogonal training symbols sent from each UE, while maintaining simultaneous data transmission in the DL.
    \item Using UL/DL reciprocity, we obtain the DL channel estimate at $(i\!-\!1)$th time slot from the UL channel estimate.
    \item Based on the DL channel estimate, we derive the DL precoder $\mathbf{V}_{b}[i]$ to process all the data streams at the $i$th time slot.
\end{enumerate}
In Fig. \ref{fig: trans_protocol}(b), the FD sequential beamforming approach of \cite{du2016sequential} is depicted, where the UEs access the UL channel in TDMA manner, and the BS initiates DL data transmission to the UEs with available CSI, instead of waiting for all the UEs to finish training. Figure \ref{fig: trans_protocol}(c) represents the conventional HD MIMO scheme, where a fraction of the total channel uses is dedicated for UL training and the rest of each time slot is utilized for data transmission in DL direction.

\subsection{Channel Estimation Error and Average DL Rate}\label{ssec: CSI Error}
To estimate the UL channel at $(i\!-\!1)$th time slot, the $K$ users transmit orthogonal training symbols $\mathbf{S}_{K}[i-1]\in \mathbb{C}^{K\times T}$, such that $\mathbb{E}[\mathbf{S}_{K}[i-1]\mathbf{S}_{K}^{\rm H}[i-1]]= TP_{K}$. Here, we utilize $T$ training symbols in the UL direction for estimation, while maintaining DL data transmission enabled by FD. Therefore, the received training symbols at node $b$ after A/D cancellation, $\mathbf{Y}_{b}[i-1]\in \mathbb{C}^{N_b\times T}$, can be expressed as
\begin{align}\label{eq: trainB}
        \nonumber\mathbf{Y}_{b}[i\!-\!1] &\triangleq \mathbf{H}^{\rm T}[i\!-\!1]\mathbf{S}_{K}[i\!-\!1] + \widetilde{\mathbf{H}}_{b,b}[i\!-\!1]\mathbf{V}_{b}[i\!-\!1]\mathbf{S}_{b}[i\!-\!1]\\
        & \qquad + \mathbf{N}_b[i\!-\!1],
\end{align}
where $\mathbf{H}^{\rm T}[i-1], \widetilde{\mathbf{H}}_{b,b}[i-1], \mathbf{V}_{b}[i-1], \mathbf{S}_{b}[i-1]$, and $\mathbf{N}_b[i-1]$ represent the UL channel, residual SI channel, DL precoder, transmit signal matrix, and AWGN noise matrix of node $b$ RXs at $(i\!-\!1)$th time slot. 

The considered MIMO channels $\mathbf{H}[i\!-\!1]$ and $\mathbf{H}[i]$ of the $(i\!-\!1)$ and $i$th time slots, respectively, are assumed temporally correlated. Using the classical Jakes model, the correlation coefficient between DL channels $\mathbf{H}[i\!-\!1]$ and $\mathbf{H}[i]$ is defined as \cite[Eq. (2.58)]{tse2005fundamentals}
\begin{equation}
    \begin{split}
        \rho \triangleq \mathbb{E}[\mathbf{H}^{\rm H}[i]\mathbf{H}[i\!-\!1]] = J_{0} (2\pi f_{d} T_{c}),
    \end{split}
\end{equation}
where $f_{d}$ is the Doppler frequency, $T_{c}$ is the time difference between two consecutive time slots, $J_0(\cdot)$ is the zero-th order Bessel function of the first kind. Using the Gauss-Markov error model, the time correlated DL channels in successive time slots can be expressed as \cite{kobayashi2011training}
\begin{equation}\label{eq: Tcorr}
    \begin{split}
        \mathbf{H}[i\!-\!1] \triangleq \rho\mathbf{H}[i] + \sqrt{1-\rho^2} \mathbf{E}[i],
    \end{split}
\end{equation}
where $\mathbf{E}[i]\in \mathbb{C}^{K\times N_b}$ is the error matrix, whose entries are modeled as IID $\mathcal{C}\mathcal{N}(0,l_{K})$, and is independent of $\mathbf{H}[i]$. Due to reciprocity, similar time correlation, as in \eqref{eq: Tcorr}, hold for all UL channels. Under this model, using UL/DL reciprocity, the received training signals at BS node $b$ at the $(i\!-\!1)$th time slot  can be written using \eqref{eq: trainB} as 
\begin{equation}\label{eq: trainB1}
    \begin{split}
        \mathbf{Y}_{b}[i\!-\!1] \triangleq& \rho\mathbf{H}^{\rm T}[i]\mathbf{S}_{K}[i\!-\!1] + \sqrt{(1-\rho^2)P_{K}} \mathbf{E}_{\text{UL}}[i]\\ &+\widetilde{\mathbf{H}}_{b,b}[i\!-\!1]\mathbf{V}_{b}[i\!-\!1]\mathbf{S}_{b}[i\!-\!1]+ \mathbf{N}_b[i\!-\!1],
    \end{split}
\end{equation}
where $\mathbf{E}_{\text{UL}}[i]\in \mathbb{C}^{N_b\times T}$ is the UL CSI delay error matrix modeled as IID $\mathcal{C}\mathcal{N}(0,l_{K})$ and independent of $\mathbf{N}_b[i\!-\!1]$. Based on \eqref{eq: trainB1}, the UL channel $\mathbf{H}^{\rm T}[i]$ can be estimated using the Minimum Mean Squared Error (MMSE) estimator as \cite{kobayashi2011training}
\begin{equation}\label{eq: chanEst}
    \begin{split}
        \widehat{\mathbf{H}}^{\rm T}[i] \triangleq&\frac{\rho \mathbf{Y}_{b}[i\!-\!1]\mathbf{S}^{\rm H}_{K}[i\!-\!1]}{\sigma_b^2 + \sigma_{r,b}^2+ (1-\rho^2)P_{K}l_{K} + \rho^2TP_{k}},
    \end{split}
\end{equation}
where $\sigma_{r,b}^2 = \|\widetilde{\mathbf{H}}_{b,b}[i\!-\!1]\mathbf{V}_{b}[i\!-\!1]\mathbf{S}_{b}[i\!-\!1]\|^2$ is the residual SI power after A/D cancellation at node $b$. From the UL estimate at the $i$th time slot, the relationship between the actual and estimated DL channels at the $i$th time slot, using the Gauss-Markov error model, can be written as\cite{iimori2019mimo}
\begin{equation}
    \begin{split}
        \mathbf{H}[i] \triangleq \sqrt{1-\tau^2_{\text{DL}}} \widehat{\mathbf{H}}[i] + \tau_{\text{DL}} \mathbf{E}_{\text{DL}}[i],
    \end{split}
\end{equation}
where $\mathbf{E}_{\text{DL}}[i]$ is the DL estimation error matrix having IID elements each modeled as $\mathcal{C}\mathcal{N}(0,l_{K})$, and $\tau_{\text{DL}}\in[0,1]$ is the Gauss-Markov error parameter that depends on the effective DL Signal-to-Noise Ratio (SNR). The case $\tau_{\rm DL}=0$ implies ideal DL CSI, whereas $\tau_{\rm DL} = 1$ signifies unavailable channel estimation. The Mean Squared Error (MSE) of the DL channel from \eqref{eq: chanEst} at the BS node $b$ is given by \cite{kobayashi2011training}
\begin{equation}\label{Eq: sig5}
    \begin{split}
        {\rm MSE_{FD}}\triangleq\tau_{\text{DL}}^2= \frac{\sigma_b^2 + \sigma_{r,b}^2+ (1-\rho^2)P_{K}l_{K}}{\sigma_b^2 + \sigma_{r,b}^2+ (1-\rho^2)P_{K}l_{K} + \rho^2TP_{k}},
    \end{split}
\end{equation}
Using the DL estimate $\widehat{\mathbf{H}}[i]$, the achievable DL rate per channel use for the proposed SCDC scheme at the $i$th time slot can be derived as
\begin{align}\label{Eq: achievable_rate_FD_1}
        \nonumber\mathcal{R}_{\text{DL}}[i]\!\triangleq& \log_2\Big(\text{det}\Big(\mathbf{I}_{K}+(1-\tau_{\text{DL}}^2){P}_{b}\widehat{\mathbf{H}}[i]\mathbf{V}_b[i]\\
        &\qquad\qquad\times\mathbf{V}_b^{\rm H}[i]\widehat{\mathbf{H}}^{\rm H}[i]\boldsymbol{\Sigma}_{k}[i]^{-1}\Big)\Big),
\end{align}
where the interference-plus-noise covariance matrix:
\begin{align}
        \nonumber\boldsymbol{\Sigma}_{k}[i] \triangleq& \sigma^2_{K}\mathbf{I}_{K}\!+\! \widetilde{\mathbf{H}}_{K,K}[i]\mathbf{s}_{K}[i]\mathbf{s}_{K}^{\rm H}[i]\widetilde{\mathbf{H}}_{K,K}^{\rm H}[i]
        \!+\! \mathbf{H}_{\text{IN}}[i]\mathbf{s}_{K}[i]\\
        &\times\mathbf{s}_{K}^{\rm H}[i]\mathbf{H}_{\text{IN}}^{\rm H}[i] + \tau_{\text{DL}}^2 {P}_{b}l_{K}\mathbf{V}_b[i]\mathbf{V}_b^{\rm H}[i].
\end{align}
\vspace*{-0.5cm}
\section{Proposed Joint Optimization Framework}
\begin{algorithm}[!t]
    \caption{Proposed FD MIMO Design}
    \label{alg:the_alg}
    \begin{algorithmic}[1]
        \renewcommand{\algorithmicrequire}{\textbf{Input:}}
       \renewcommand{\algorithmicensure}{\textbf{Output:}}
        \REQUIRE $\widehat{\mathbf{H}}_{b,b}$, $\widehat{\mathbf{H}}$, $\widehat{\mathbf{H}}_{K,K}$, ${P}_b$, ${P}_{K}$, $N$, and $N_b$.
        \ENSURE $\mathbf{V}_b$, $\mathbf{C}_b$, $\mathbf{C}_{K}$,  $\mathbf{D}_b$, and $\mathbf{D}_{K}$.
        \STATE Obtain the $N$-tap analog canceller $\mathbf{C}_b$ using \cite[Eq. 1]{alexandropoulos2017joint}.
        \STATE Obtain $\mathbf{Q}_b$ including the $N_b$ right-singular vectors of $(\widehat{\mathbf{H}}_{b,b} + \mathbf{C}_b)$ corresponding to the singular values in descending order.
        \STATE Set $\mathbf{C}_{K}= -\widehat{\mathbf{H}}_{K,K}$.
        \FOR{$\alpha=N_b, N_b-1, \dots, 2$}
            \STATE Set $\mathbf{F}_b=[\mathbf{Q}_b]_{(:,N_b-\alpha+1 : N_b)}$.
            \STATE Set $\mathbf{W}_b = \text{sort}(\widehat{\mathbf{H}}\mathbf{F}_b)$, rearranging the rows of $\widehat{\mathbf{H}}\mathbf{F}_b$ based on descending row vector norm.
            \STATE Set $\mathbf{Z}_b = [\mathbf{W}_b]_{(1:m_b,:)}$.
            \STATE Set $\mathbf{G}_b = \beta_{\text{ZF}} \mathbf{Z}_b^{\rm H}(\mathbf{Z}_b\mathbf{Z}_b^{\rm H})^{-1}$, where $\beta_{\text{ZF}}$ is the normalization constant that ensures $\mathbb{E}[\text{tr}(\mathbf{G}_b\mathbf{G}_b^{\rm H})]= 1$.
            \STATE Set the DL precoder as $\mathbf{V}_b= \mathbf{F}_b \mathbf{G}_b$.
            \IF {${P}_{b}\|[(\widehat{\mathbf{H}}_{b,b}+\mathbf{C}_{b})\mathbf{V}_b]_{(j,:)}\|^2\!\leq\! \lambda_b, \forall{j\!=\!1,2,\ldots,N_b}$, and $P_{K}\|[(\widehat{\mathbf{H}}_{k,k}+\mathbf{C}_{k})]_{(j,:)}\|^2\leq \lambda_k, \,\forall j\!=\!1,2,\ldots,K$}
            \STATE Output $\mathbf{V}_b$, $\mathbf{C}_b$, $\mathbf{C}_{K}$, $\mathbf{D}_b=-(\widehat{\mathbf{H}}_{b,b}+ \mathbf{C}_b)$, $\mathbf{D}_K=-(\widehat{\mathbf{H}}_{K,K}+ \mathbf{C}_K)$, and stop the algorithm.
            \ENDIF
        \ENDFOR
        \STATE Set $\mathbf{V}_b=[\mathbf{Q}_b]_{(:,N_b)}$.
        \IF {${P}_{b}\|[(\widehat{\mathbf{H}}_{b,b}+\mathbf{C}_{b})\mathbf{V}_b]_{(j,:)}\|^2\leq \lambda_b, \forall {j}\!=\!1,2,\ldots,N_b$, and $P_{K}\|[(\widehat{\mathbf{H}}_{k,k}+\mathbf{C}_{k})]_{(j,:)}\|^2\leq \lambda_k, \,\forall j\!=\!1,2,\ldots,K$} 
            \STATE Output $\mathbf{V}_b$, $\mathbf{C}_b$, $\mathbf{C}_{K}$, $\mathbf{D}_b=-(\widehat{\mathbf{H}}_{b,b}+ \mathbf{C}_b)$, $\mathbf{D}_K=-(\widehat{\mathbf{H}}_{K,K}+ \mathbf{C}_K)$, and stop the algorithm.
        \ELSE
        \STATE Output that the $\mathbf{C}_b$ realizations or $\mathbf{C}_K$ do not meet the receive RF saturation constraints.
        \ENDIF
    \end{algorithmic}
\end{algorithm}
In this section, we focus on the joint design of the digital TX precoder $\mathbf{V}_b$, the analog SI cancellers $\mathbf{C}_b$ and ${c}_k, \forall k=1,\cdots,K$, as well as the digital cancellers $\mathbf{D}_b$ and ${d}_k, \forall k$ at BS node $b$ and $K$ UE nodes, respectively, maximizing the estimated achievable DL rate of $i$th time slot. It is to be noted that, for simplicity, we omit the time slot index in the subsequent optimization problem, as we are only dealing with variables at $i$th time slot, unless mentioned otherwise. Based on the estimated DL channel $\widehat{\mathbf{H}}$, SI channels $\widehat{\mathbf{H}}_{b,b}$ and $\widehat{\mathbf{H}}_{K,K}$ at BS and UE nodes, respectively, the considered optimization problem is expressed as: 
\begin{align}\label{eq: optimization_eq}
        \nonumber\underset{\substack{\mathbf{V}_b,\mathbf{C}_b, \mathbf{D}_{b}\\ \mathbf{C}_{K}, \mathbf{D}_{K}}}{\text{max}} &\log_2\!\left(\!\text{det}\left(\mathbf{I}_{K}\!+\!(1\!-\!\tau_{\text{DL}}^2){P}_{b}\widehat{\mathbf{H}}\mathbf{V}_b\mathbf{V}_b^{\rm H}\widehat{\mathbf{H}}^{\rm H}\boldsymbol{\Sigma}_{k}^{-1}\right)\right)\\
        \text{\text{s}.\text{t}.}\quad
        &P_b\|[(\widehat{\mathbf{H}}_{b,b}+\mathbf{C}_{b})\mathbf{V}_b ]_{(j,:)}\|^2\leq \lambda_b \,\forall j=1,2,\ldots,N_b,\nonumber\\
        &P_{K}\|[(\widehat{\mathbf{H}}_{k,k}+\mathbf{C}_{k})]_{(j,:)}\|^2\leq \lambda_k, \,\forall j=1,2,\ldots,K,\nonumber\\
        &\mathbb{E}\{\|\mathbf{V}_b \mathbf{s}_{b}\|^2\}\leq {P}_{b},\,\,\text{and}\,\,\,
        \mathbb{E}\{\|\mathbf{s}_{K}\|^2\}\leq {P}_{K}.
\end{align}
In this formulation, the first constraint imposes the RX RF chain saturation threshold $\lambda_{b}$ after analog cancellation at BS node $b$. As previously discussed, this threshold ensures proper reception of the training symbols by all $N_b$ RX RF chains of the BS, which means that the UL channel can be efficiently estimated using \eqref{eq: chanEst}. The second constraint enforces the saturation threshold $\lambda_{K}$ at the $K$ UEs assuring feasible decoding of BS's information data symbols. The final two constraints in \eqref{eq: optimization_eq} refer to the nodes' average transmit powers.

The optimization problem in \eqref{eq: optimization_eq} is quite difficult to tackle, since it is non-convex including couplings among the optimization variables. In this paper, we suboptimally solve it using an alternating optimization approach, leaving other possibilities for future work. To this end, we start with an allowable $\mathbf{C}_{b}$ realization given the available number of analog canceller taps $N$, where the tap values are set to be the respective amplitude elements of the estimated SI channel $\widehat{\mathbf{H}}_{b,b}$. Based on the chosen $\mathbf{C}_{b}$, we seek for the precoding matrix $\mathbf{V}_b$ maximizing the DL rate, while meeting the first constraint for the BS analog SI cancellation threshold $\lambda_{b}$. This procedure is repeated for all allowable realizations of $\mathbf{C}_{b}$ to find the best pair of $\mathbf{C}_{b}$ and $\mathbf{V}_b$. Adopting the approach in \cite{alexandropoulos2017joint}, the BS precoder for DL data communication is constructed as $\mathbf{V}_b= \mathbf{F}_b \mathbf{G}_b$, where $\mathbf{F}_b \in \mathbb{C}^{N_b \times \alpha}$ aims at reducing the residual SI after analog cancellation and $\mathbf{G}_b \in \mathbb{C}^{\alpha \times m_b}$ is the Zero-Forcing (ZF) beamformer maximizing the rate of the effective DL channel $\widehat{\mathbf{H}}\mathbf{F}_b$. The parameter $\alpha$ is a positive integer taking the values $1 \leq \alpha \leq N_b$. At each of the UE nodes, a single-tap analog canceller is employed resulting in a diagonal analog SI canceller matrix $\mathbf{C}_{K}= -\widehat{\mathbf{H}}_{K,K}$. To maximize the signal-to-interference-plus-noise ratio, the residual SI is further reduced by setting the digital cancellation signal at both FD nodes as their respective complementary residual SI channels after analog SI cancellation. For each allowable realization of $\mathbf{C}_{b}$, the proposed solution for the considered optimization problem \eqref{eq: optimization_eq} is summarized in Algorithm \ref{alg:the_alg}.

\begin{figure}[!tpb]
\centering
    \includegraphics[width=\linewidth]{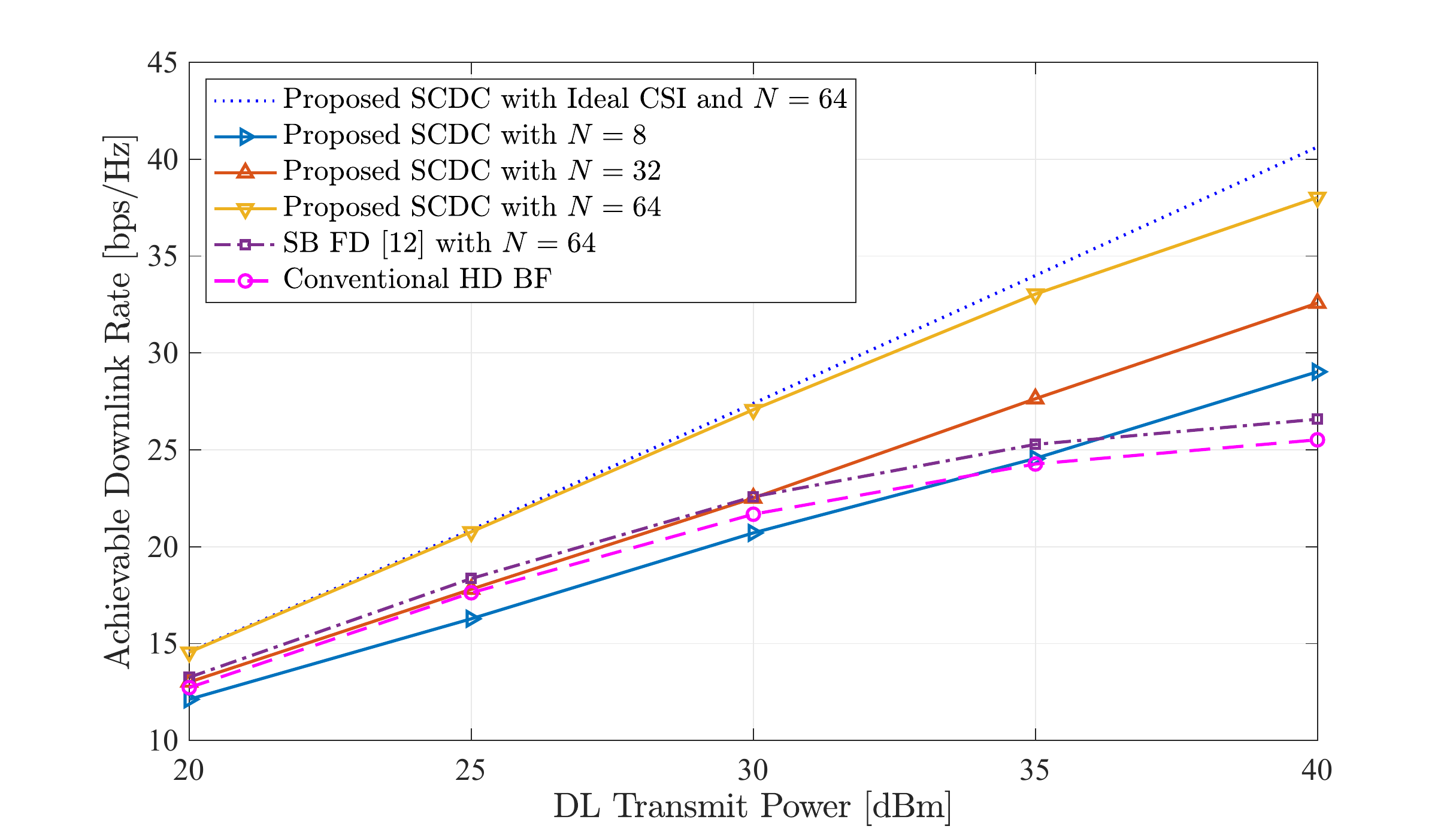}
	\caption{\footnotesize Achievable DL Rate w.r.t DL transmit power for $N_b=8$, $K=4$, $f_d =50$Hz and different analog canceller taps $N$.}
	\label{fig: DL_Power}
\end{figure}
\section{Simulation Results and Discussion}
In this section, we provide representative simulation results of the proposed FD-based MU MIMO simultaneous data communication and channel estimation approach. In Section \ref{ssec: ComparedDesigns}, we describe the existing FD and HD MIMO solutions to be compared with our proposed approach. The simulations parameters and assumptions are detailed in Section \ref{ssec: Sim_param}, whereas the hardware complexity, and achievable DL rate results are presented in Section \ref{ssec: RateResults}.

\subsection{Compared FD and HD MIMO Designs}\label{ssec: ComparedDesigns}
We compare our novel FD SCDC scheme with the Sequential Beamforming (SB) FD approach presented in \cite{du2016sequential}, where the UEs send their training symbols using TDMA. Contrary to the conventional HD MIMO approach, which is also illustrated here for comparison, in SB FD approach, the BS node does not wait for all the UEs to finish training. In SB FD, the BS starts transmitting to the UE with available CSI at the BS node, while receiving training from other UEs. The SB FD approach considers perfect analog SI cancellation, which includes full-tap analog SI canceller. In addition, we also illustrate the proposed FD approach with ideal CSI and full-tap analog SI canceller.
\subsection{Simulation Parameters}\label{ssec: Sim_param}
We perform an extensive simulation following the FD MIMO architecture illustrated in Fig$.$~\ref{fig: transmission}. We have considered an $8\times8$ (i.e., $N_b=8$) FD MIMO BS node $b$ serving $K=4$ single-antenna FD UE nodes. DL, UL, and inter-node interference channels are assumed as block Rayleigh fading channels with a pathloss of $110$dB. The SI channels at the BS and all UE nodes are simulated as Rician fading channels with a $\kappa$-factor of $30$dB and pathloss of $40$dB \cite{duarte2012experiment}. We have considered a narrowband communication system with a bandwidth and carrier frequency of $1.4$MHz and $2.4$GHz, respectively. RX noise floors at all nodes were assumed to be $-100$dBm. To this end, the RXs have effective dynamic range of $50$dB provided by the $12$-bit analog-to-digital converters (ADC) for a Peak-to-Average-Power-Ratio (PAPR) of $10$ dB \cite{AD3241}. Therefore, the residual SI power after analog SI cancellation at the input of each RX chain has to be below $-50$dBm to avoid RX RF chain saturation. Furthermore, non-ideal multi-tap analog canceller is considered with steps of $0.02$dB for attenuation and $0.13^\circ$ for phase as in \cite{alexandropoulos2017joint}. All the user nodes employ single-tap SI canceller, where the MIMO BS node requires an $N$-tap analog SI canceller. We have used $1000$ independent Monte Carlo simulation runs to calculate the performance of all considered designs. Every transmission time slot is considered to be $1$ms with $T=400$ symbols. For the compared HD and SB FD scheme, $10\%$ of the total symbols were dedicated for UL channel sounding. The UL transmit power of each UE node is limited to $10$dBm for all considered designs. 
\begin{figure}[!tpb]
\centering
    \includegraphics[width=\linewidth]{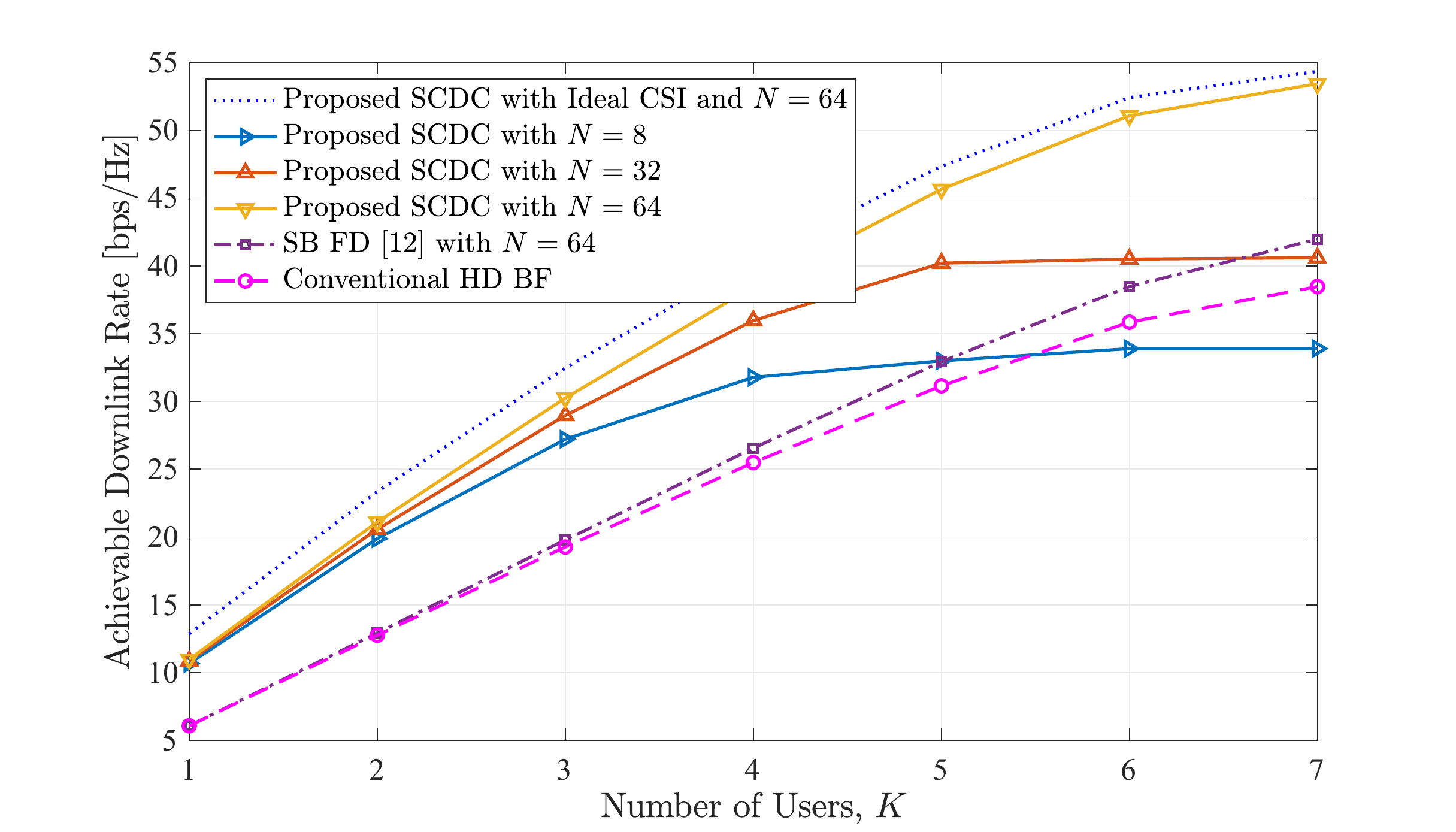}
	\caption{\footnotesize Achievable DL Rate w.r.t number of UEs $K$ for $N_b=8$, $P_b=40$dBm, $f_d =50$Hz and different analog canceller taps $N$.}
	\label{fig: DL_Users}
\end{figure}
\subsection{Analog Canceller Complexity and Achievable DL Rate}\label{ssec: RateResults}
We illustrate the DL rate performance of the proposed and other considered designs in Fig. \ref{fig: DL_Power} with respect to DL transmit power for $K=4$, $f_d =50$Hz and varying analog canceller taps $N$. The proposed SCDC scheme with full-tap analog canceller with $N=64$ provides DL rate close to the ideal case for transmit power below $30$dBm and outperforms all other considered designs for all transmit powers. After $50\%$ reduction of analog canceller taps, the proposed approach with $N=32$ still provides $1.25\times$ DL rate compared to the SB FD \cite{du2016sequential} and conventional HD approach for transmit power of $40$dBm, while achieving identical DL rate performance for low transmit powers. Furthermore, we consider a more extreme case with $N=8$, which results in an analog SI canceller with only $12.5\%$ taps. Although, such low-complexity SCDC approach provides less DL rate compared to the SB FD approach, it can achieve higher performance for high transmit powers above $35$dBm.

In Fig. \ref{fig: DL_Users}, we plotted the achievable DL rate of the considered transmission schemes with respect to the the number of UEs $K$ being served by the FD BS node for a fixed DL transmit power of $40$dBm. It is evident from the figure that, the proposed SCDC scheme with $N=64$ provides substantial DL rate increment compared to the SB FD \cite{du2016sequential} and the conventional FD design for any number of UEs. However, the proposed approach with $N=32$ can efficiently serve $6$ UEs providing better DL rate than the SB FD approach. For a further reduction of canceller taps, the proposed SCDC approach with $N=8$ achieves higher rate than SB FD and conventional HD scheme up to $5$ UEs. Therefore, the proposed SCDC scheme provides a flexible design with an important trade-off between number of UEs and analog SI canceller taps that achieves higher DL rate performance compared to the full-tap SB FD approach.

\begin{figure}[!tpb]
\centering
    \includegraphics[width=\linewidth]{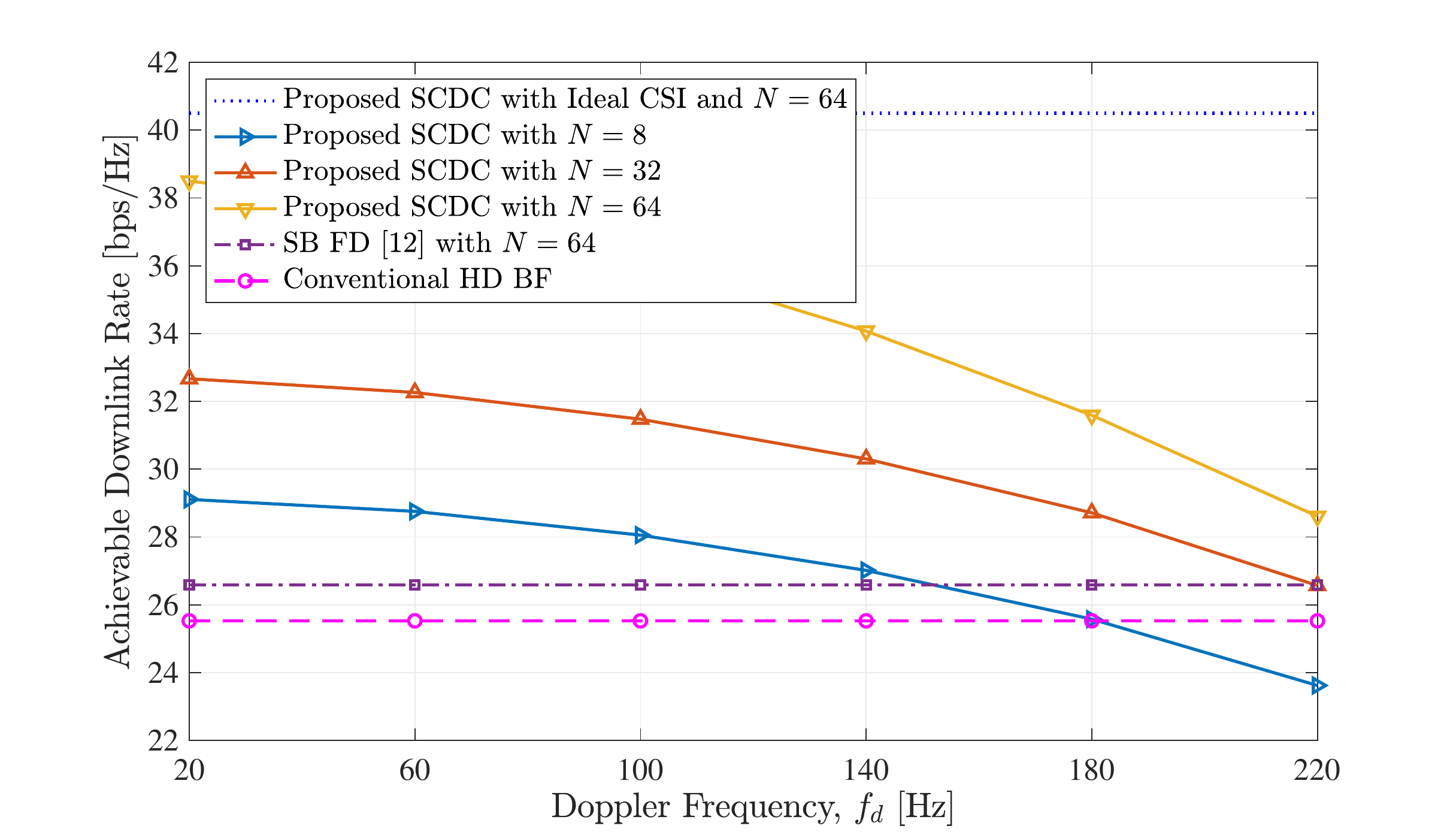}
	\caption{\footnotesize Achievable DL Rate w.r.t Doppler frequency $f_d$ for $N_b=8$, $P_b=40$dBm, $K=4$ and different analog canceller taps $N$.}
	\label{fig: DL_Doppler}
\end{figure}

We showcase the DL rate performance of the considered design with respect to the Doppler frequency $f_d$ in Fig. \ref{fig: DL_Doppler} for $K=4$ UEs and fixed DL transmit power of $40$dBm. As described in \ref{ssec: CSI Error}, the successive time slots are correlated and the correlation coefficient is measured using $f_d$ given $T_c = 1$ms. It is evident from Fig. \ref{fig: DL_Doppler} that the increment of $f_d$ only affects the proposed scheme, as it estimates the DL channel based on the training symbols acquired in the preceding time slot. However, it is shown that the proposed schemes with $N=64$ and $N=32$ taps provide superior rate compared to the SB FD\cite{du2016sequential} and the conventional HD approach for a high Doppler frequency of $220$Hz, which for our considered communication parameter represents a relative velocity $100$km/h between BS and UE nodes. The proposed SCDC scheme with $N=8$ can achieve higher rate for $f_d$ up to $140$Hz. Therefore, we conclude that, compared to the SB FD and conventional HD scheme, the proposed SCDC approach can achieve higher rates with $50\%$ less taps. 

\section{Conclusion}
In this paper, we proposed a multi-user FD MIMO communication system for simultaneous DL information data transmission and UL CSI estimation with reduced complexity multi-tap analog SI cancellation. Considering an MMSE-based channel estimation error model, we presented a unified optimization framework for the joint design of digital TX precoding and A/D SI cancellation. Our performance evaluation results demonstrated that the proposed SCDC protocol is capable of achieving improved achievable DL rates compared to the existing FD and conventional HD systems, requiring reduced complexity analog cancellation compared to conventional FD MIMO architectures.

\squeezeupann
\squeezeupann
\section*{Acknowledgments}
This work was partially funded by the National Science Foundation CAREER award \#1620902.

\squeezeupann

\bibliographystyle{IEEEtran}
\bibliography{IEEEabrv,ms}

\begin{thebibliography}{10}
\providecommand{\url}[1]{#1}
\csname url@samestyle\endcsname
\providecommand{\newblock}{\relax}
\providecommand{\bibinfo}[2]{#2}
\providecommand{\BIBentrySTDinterwordspacing}{\spaceskip=0pt\relax}
\providecommand{\BIBentryALTinterwordstretchfactor}{4}
\providecommand{\BIBentryALTinterwordspacing}{\spaceskip=\fontdimen2\font plus
\BIBentryALTinterwordstretchfactor\fontdimen3\font minus
  \fontdimen4\font\relax}
\providecommand{\BIBforeignlanguage}[2]{{%
\expandafter\ifx\csname l@#1\endcsname\relax
\typeout{** WARNING: IEEEtran.bst: No hyphenation pattern has been}%
\typeout{** loaded for the language `#1'. Using the pattern for}%
\typeout{** the default language instead.}%
\else
\language=\csname l@#1\endcsname
\fi
#2}}
\providecommand{\BIBdecl}{\relax}
\BIBdecl

\bibitem{sabharwal2014band}
A.~Sabharwal, P.~Schniter, D.~Guo, D.~W. Bliss, S.~Rangarajan, and R.~Wichman,
  ``In-band full-duplex wireless: Challenges and opportunities.'' \emph{IEEE J.
  Sel. Areas Commun.}, vol.~32, no.~9, pp. 1637--1652, Sep. 2014.

\bibitem{duarte2012experiment}
M.~Duarte, C.~Dick, and A.~Sabharwal, ``Experiment-driven characterization of
  full-duplex wireless systems,'' \emph{IEEE Trans. Wireless Commun.}, vol.~11,
  no.~12, pp. 4296--4307, Dec. 2012.

\bibitem{bharadia2013full}
D.~Bharadia, E.~McMilin, and S.~Katti, ``Full duplex radios,'' in \emph{Proc.
  ACM SIGCOMM}, Hong Kong, China, 12-16 Aug. 2013, pp. 375--386.

\bibitem{smida2017reflectfx}
B.~Smida and S.~Khaledian, ``Reflect{FX}: In-band full-duplex wireless
  communication by means of reflected power,'' \emph{IEEE Trans. Commun.},
  vol.~65, no.~5, pp. 2207--2219, May 2017.

\bibitem{islam2019comprehensive}
M.~A. Islam and B.~Smida, ``A comprehensive self-interference model for
  single-antenna full-duplex communication systems,'' in \emph{Proc. IEEE ICC},
  Shanghai, China, May 2019, pp. 1--7.

\bibitem{alexandropoulos2020full}
G.~C. Alexandropoulos, M.~A. Islam, and B.~Smida, ``Full duplex hybrid {A/D}
  beamforming with reduced complexity multi-tap analog cancellation,'' in
  \emph{Proc. {IEEE SPAWC}}, Atlanta, USA, May 2020, pp. 1--6.

\bibitem{islam2019unified}
M.~A. Islam, G.~C. Alexandropoulos, and B.~Smida, ``A unified beamforming and
  {A/D} self-interference cancellation design for full duplex {MIMO} radios,''
  in \emph{Proc. IEEE PIMRC}, Istanbul, Turkey, Sep. 2019, pp. 1--7.

\bibitem{riihonen2011mitigation}
T.~Riihonen, S.~Werner, and R.~Wichman, ``Mitigation of loopback
  self-interference in full-duplex {MIMO} relays,'' \emph{IEEE Trans. Signal
  Process.}, vol.~59, no.~12, pp. 5983--5993, Dec. 2011.

\bibitem{everett2016softnull}
E.~Everett, C.~Shepard, L.~Zhong, and A.~Sabharwal, ``Softnull: Many-antenna
  full-duplex wireless via digital beamforming,'' \emph{IEEE Trans. Wireless
  Commun.}, vol.~15, no.~12, pp. 8077--8092, Dec. 2016.

\bibitem{alexandropoulos2017joint}
G.~C. Alexandropoulos and M.~Duarte, ``Joint design of multi-tap analog
  cancellation and digital beamforming for reduced complexity full duplex
  {MIMO} systems,'' in \emph{Proc. IEEE ICC}, Paris, France, May 2017, pp.
  1--7.

\bibitem{du2015mu}
X.~Du, J.~Tadrous, C.~Dick, and A.~Sabharwal, ``{MU-MIMO} beamforming with
  full-duplex open-loop training,'' in \emph{Proc. IEEE SPAWC}, Stockholm,
  Sweden, Jun. 2015, pp. 301--305.

\bibitem{du2016sequential}
X.~Du, J.~Tadrous, and A.~Sabharwal, ``Sequential beamforming for multi-user
  {MIMO} with full-duplex training,'' \emph{IEEE Trans. Wireless Commun.},
  vol.~15, no.~12, pp. 8551--8564, Dec. 2016.

\bibitem{mirza2018performance}
J.~Mirza, G.~Zheng, K.-K. Wong, S.~Lambotharan, and L.~Hanzo, ``On the
  performance of multi-user {MIMO} systems relying on full-duplex {CSI}
  acquisition,'' \emph{IEEE Trans. Wireless Commun.}, vol.~66, no.~10, pp.
  4563--4577, Oct. 2018.

\bibitem{islam2020simultaneous}
M.~A. Islam, G.~C. Alexandropoulos, and B.~Smida, ``Simultaneous downlink data
  transmission and uplink channel estimation with reduced complexity full
  duplex {MIMO} radios,'' in \emph{Proc. {IEEE ICC}}, Dublin, Ireland, Jun.
  2020, pp. 1--6.

\bibitem{kobayashi2011training}
M.~Kobayashi, N.~Jindal, and G.~Caire, ``Training and feedback optimization for
  multi-user {MIMO} downlink,'' \emph{IEEE Tran. on Commun.}, vol.~59, no.~8,
  pp. 2228--2240, May 2011.

\bibitem{tse2005fundamentals}
D.~Tse and P.~Viswanath, \emph{Fundamentals of wireless communication}.\hskip
  1em plus 0.5em minus 0.4em\relax Cambridge university press, 2005.

\bibitem{iimori2019mimo}
H.~Iimori, G.~T.~F. de~Abreu, and G.~C. Alexandropoulos, ``{MIMO} beamforming
  schemes for hybrid {SIC} {FD} radios with imperfect hardware and {CSI},''
  \emph{IEEE Trans. Wireless Commun.}, vol.~18, no.~10, pp. 4816--4830, Oct.
  2019.

\bibitem{AD3241}
\emph{Analog-to-Digital Converters}, ADC3241, Texas Instruments, 2016.

\end{thebibliography}
\end{document}